# Experimental study of THGEM detector with mini-rim*

ZHANG Ai-Wu(章爱武)[1;1)]   YU Bo-Xiang(俞伯祥)[1;2)]   XIE Yu-Guang(谢宇广)[1]   LIU Hong-Bang(刘宏邦)[2]
AN Zheng-Hua(安正华)[1]   WANG Zhi-Gang(王志刚)[1]   CAI Xiao(蔡啸)[1]   SUN Xi-Lei(孙希磊)[1]
SHI Feng(石峰)[1]   FANG Jian(方建)[1]   XUE Zhen(薛镇)[1]   LU Qi-Wen(吕绮雯)[1,3]
SUN Li-Jun(孙丽君)[1]   GE Yong-Shuai(葛永帅)[1]   LIU Ying-Biao(刘颖彪)[1]
HU Tao(胡涛)[1]   ZHOU Li(周莉)[1]   LU Jun-Guang(吕军光)[1]

1 (Institute of High Energy Physics, Chinese Academy of Sciences, Beijing 100049, China)
2 (Graduate University of Chinese Academy of Sciences, Beijing 100039, China)
3 (Shanxi University, Taiyuan 030006, China)

**Abstract**   The gas gain and energy resolution of single and double THGEM detectors ($5\times5\text{cm}^2$ effective area) with mini-rims (rim is less than $10\mu\text{m}$) were studied. The maximum gain can reach $5\times10^3$ and $2\times10^5$ for single and double THGEM respectively, while the energy resolution of 5.9 keV X-ray varied from 18% to 28% for both single and double THGEM detectors of different hole sizes and thicknesses. All the experiments were investigated in mixture of noble gases(argon,neon) and small content of other gases(iso-butane,methane) at atmospheric pressure.

## 1  Introduction

In the past decades, the Micro-Pattern Gases Detectors(MPGDs) have been experienced a flourishing development, and the most successful MPGDs are gaseous electron multipliers(GEM)[1, 2] and the Micromegas[3]. The development of the THick Gas Electron Multiplier (THGEM) was motivated by the need for robust large-area, fast radiation imaging detectors with moderate localization resolution[4–7]. In general, the THGEMs were manufactured with standard PCB technology by precise drilling on double-face Cu-clad FR-4 or G10 substrates, and a metal-free clearance ring, the rim, surrounding the hole was obtained by Cu etching (Fig. 1(a)). The rim enhances the THGEM's immunity to discharges, leading to higher gains compared to rimless holes[8, 9]. Conversely, large rim may be responsible for charging up and polarization of the substrate, thus long time is needed to get stable gain[10]. And without the rim, the detector is characterized by an order of magnitude lower gain than THGEMs with large rims[11, 12].

The THGEM detector has also been studied by us[13, 14]. However, there is a non-concentricity in the etching technics, which makes a bias between the center of the drilled hole and the etched circular ring. This will make the THGEM easily discharge when the high voltage is on. In addition, the cost will be much higher to improve the accuracy of ring. In the present work, the THGEMs were produced without etching the rims in the PCB factory[15], and then some chemical treatments and ultrasonic cleaning processes were taken in our laboratory. Finally we got very nice THGEMs with mini-rim, which is less than $10\mu\text{m}$. We found they have very good performances(robust,high gain,stable,etc.), and the detector cost is considerably low.

## 2  Preparation of the THGEMs

Three kinds of THGEM plates were designed in this work. The thicknesses of the plates were 0.3, 0.5 and 0.8mm respectively, while the hole diameters were kept to be the same as each thickness and the pitchs were twice of the thicknesses. The active area

Received 1 May 2011

  * Supported by Youth Found of Institute of High Energy Physics, Chinese Academy of Sciences
   1) E-mail: zhangaw@ihep.ac.cn
   2) E-mail: yubx@ihep.ac.cn (corresponding author)

was designed to be 5×5cm$^2$. In this configuration, the ratio of the hole area to the active area (which is 22.67 %) was consistent with the standard GEM detectors(70$\mu$m thickness and hole diameter,140$\mu$m pitch[2]).

The holes of the THGEM plates were drilled by the PCB manufacturer[15], and no rim was etched. The post processing includes chemical treatment and ultrasonic cleaning. Fig. 1(b) is a comparison of the hole shape before and after these processes. From this figure a mini-rim (less than 10$\mu$m) can be seen.

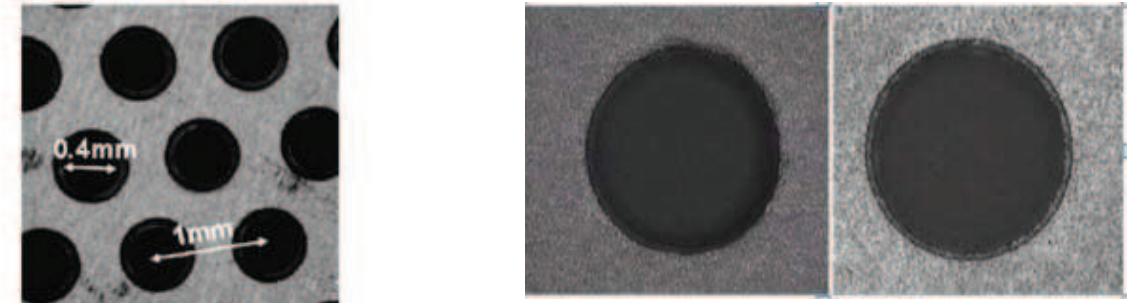

(a) The standard THGEM structure in the literatures

(b) A single hole before (left) and after (middle,rim is less than 10$\mu$m) the chemical and cleaning processes

Fig. 1. Images of the THGEM plates.

## 3  Results of the experiment

We have measured single and double THGEM plates using Fe-55 X-ray. The distance of the drift region and the induction region was set to be 5mm and 3mm, respectively, while the transfer distance for the double-THGEM was 3mm.

The gain and energy resolution were compared for the three types of plates. In addition, the Ar/ISO(95/5) and Ne/CH$_4$(95/5) were used as the operation gas. The Ortec charge sensitive preamplifier 142AH and primary amplifier 450 and Multi-Channel Analyzer TRUMP-PCI-8K (MCA) have been used as the electronic system. The system was calibrated with an Ortec pulse generator 415 and a standard capacity( 2pC/mV).

### 3.1  Single THGEM result

Fig. 2 refers to the gain of single THGEM plates of different thicknesses. From this figure one can see that the thicker of the THGEM the higher of the work voltage. The maximum gain of a single THGEM can reach about 5×10$^3$. The energy resolution, which is also shown in this figure, varies from 18% to 28%. A typical energy spectrum with a resolution 20.08% is shown in Fig. 3.

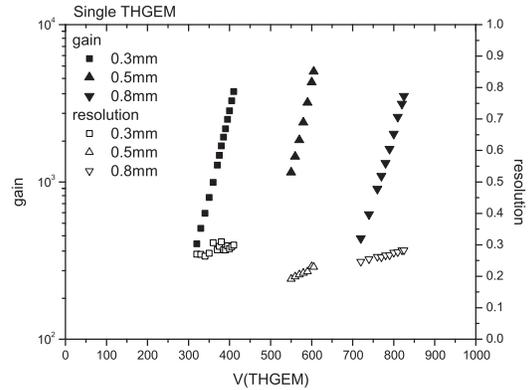

Fig. 2. The gain and energy resolution of single thgem.In Ar/ISO(95/5) mixture, drift field is

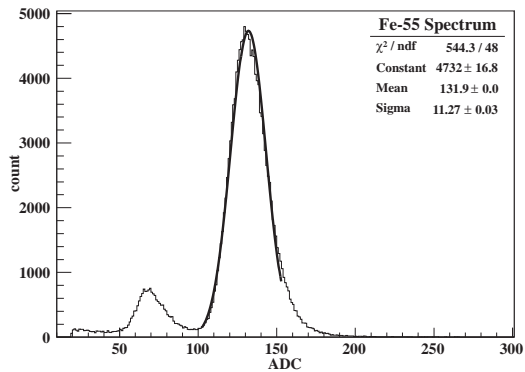

Fig. 3. A typical Fe-55 Energy Spectrum, in Ar/ISO(95/5) mixture.The energy resolution is 2.35×11.27/131.9=20.08%.

In addition, the influences of the drift and induction field on the gain and energy resolution have been investigated, and the results are shown in Fig. 4. From this figure we can see that the energy resolution becomes worse if the drift field is larger than 1kV/cm, at the same time the gain decreases. On the other hand, higher gain could be reached at higher induction field, although the resolution lose a little to some extent. It is better for an induction field not exceed 3kV/cm. As an optimization, the drift filed and the induction field were chosen to be 1kV/cm and 2.7kV/cm respectively in the following measurements.

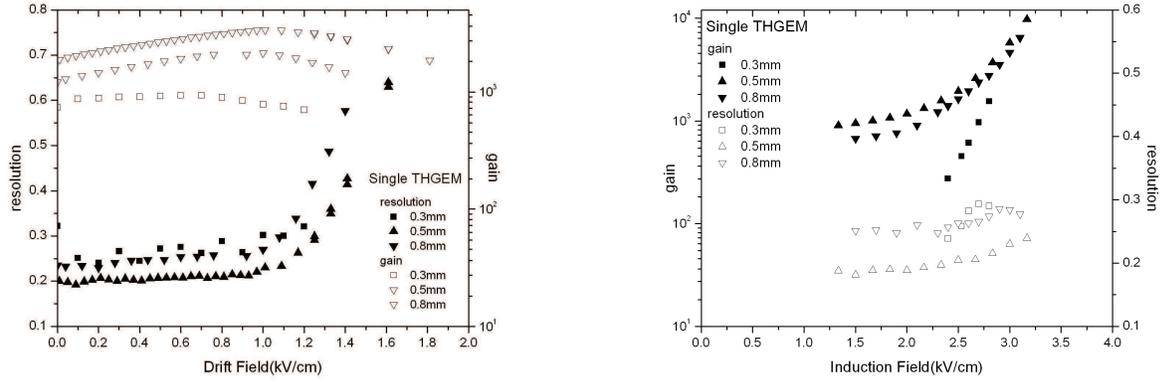

Fig. 4. The influences of drift (left) and induction (right) field to the gain and resolution, in Ar/ISO(95/5) mixture, THGEM thickness is 0.5mm and voltage is 595V.

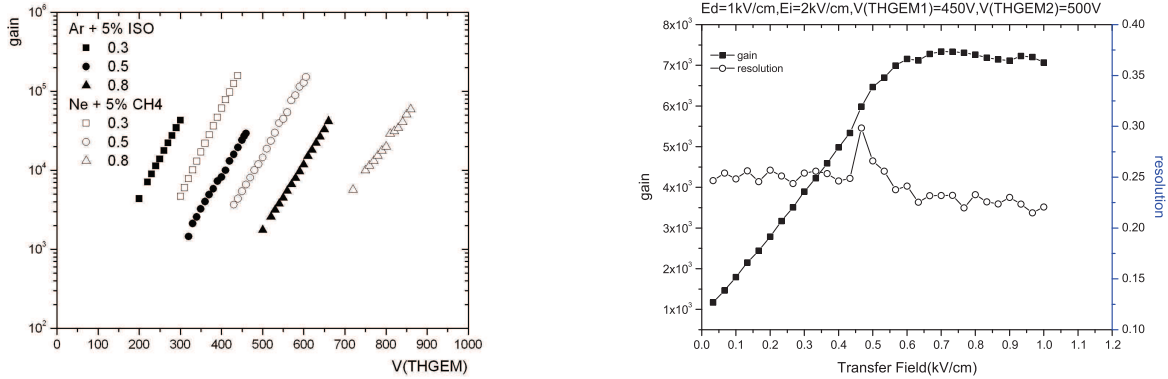

(a) Gain in different gases

(b) Influence of the transfer field. The gas is Ar/ISO(95/5), and the bad resolution at 4.5kV/cm was due to some discharges

Fig. 5. Double-THGEM results.

### 3.2 Double THGEM results

In this section, we mainly measured the gain of double-THGEM for the three thickness plates in different gases. Fig. 5(a) shows the gain of one of the two THGEM plates, while the other one was fixed at a proper voltage(the ratio of $V_{THGEM}$ to thickness were kept the same for the three thicknesses of plates). In this figure we can see the gain in argon and iso mixture is about $4 \times 10^4$ while in neon and methane gas is $2 \times 10^5$.

In addition, we studied the influence of the transfer field, which is the field between the two THGEM plates. Fig. 5(b) shows that there is a plateau after 0.6kV/cm. The energy resolution slightly goes better as the transfer field increasing.

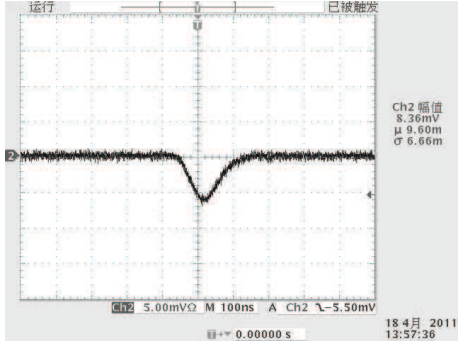

Fig. 6. A signal without amplifiers.

As the gain reached $10^5$ level, we briefly tested the double-THGEM plates without amplifiers. It was confirmed that the signals can be directly seen from an oscilloscope(with 50Ω impedance) even without both pre- and primary amplifiers. Fig. 6 is a signal waveform graph got nearby the breakdown point, from which the gain is estimated to be about $3 \times 10^5$ using the approximate formula $g = VT/2ReN$, where $V$ is the amplitude(∼6mV), $T$ is the signal width(∼100ns), $R$ is 50Ω, $e$ is the electron charge and $N$ the primary ionization number(∼166 in neon and methane mixture). This calculation is consistent with the value measured with amplifiers.

## 4 Conclusion

The THGEM detector has been researched a lot and has applications in many areas. As a supplement, the THGEMs with mini-rims have been studied and very good performances are gained. Following are some characters of our THGEMs:

- they are very robust, and will not easily be damaged by electrical discharge.

- High gain and good energy resolution can be reached, as long as the careful cleaning.

- It is very cheap to produce this kind of THGEMs (about 0.1RMB per square centimeter).The chemical treatment can be implemented easily in the laboratory.

In the future, we are going to study the THGEMs with other geometries (hole size, thickness and pitch), as well as larger area(10×10cm²). We hope more broad applications will carry out, such as photon detectors with CsI cathodes, the Dark Matter search in nobel-liquid, and so on.

**Acknowledgments** *This work is supported by Youth Fund of IHEP.We would like to express our sincere appreciation to Prof. XIE Yi-Gang for his friendly discussions and kind supports.*